\documentclass[preprint]{aastex}

\newcommand{\rr}{{\bf r}}
\newcommand{\kk}{{\bf k}}
\newcommand{\bcdot}{{\bf \cdot}}
\newcommand{\xithree}{\xi^{[3]}}
\newcommand{\wpthree}{w_p^{[3]}}
\newcommand{\wpthreen}{w_{p,n}^{[3]}}

\slugcomment{ApJ, 614:000, 2004}
\shorttitle{Projected 3-Point Correlation Functions}
\shortauthors{Zheng}

\begin{document}

\title{Projected Three-Point Correlation Functions and Galaxy Bias}
\author{Zheng Zheng}
\affil{ Department of Astronomy, Ohio State University,
        Columbus, OH 43210, USA
      }
\email{zhengz@astronomy.ohio-state.edu}

\begin{abstract}
The three-point correlation function (3PCF) can now be measured in large
galaxy redshift surveys, but in three dimensions its interpretation is
complicated by the presence of redshift-space distortions. I investigate 
the projected 3PCF, where these distortions are eliminated by integrating
over the redshift dimension, as is commonly done for the two-point correlation
function. The calculation of the projected 3PCF from the real-space, 
three-dimensional bispectrum is greatly simplified by expanding both 
quantities in Fourier components, analogous to Szapudi's expansion of 
the three-dimensional quantities in multipole components. In the weakly
nonlinear regime, the bispectrum can be well represented by the first few
Fourier components. There is a well-known relation between the reduced 
3PCFs of matter and galaxies in the weakly nonlinear regime, which can be
used to infer galaxy bias factors if the real-space three-dimensional galaxy
correlation functions (two-point and three-point) can be measured. I show
that the same relation holds for the reduced {\it projected} 3PCFs if these
are properly defined. These results should aid determinations of galaxy 
bias from large redshift surveys by eliminating the complication of 
redshift-space distortions.
\end{abstract}

\section{Introduction}

The three-point correlation function (3PCF), or its Fourier transform, 
the bispectrum, is a valuable complement to two-point statistics 
in characterizing galaxy clustering.
The behavior of the 3PCF of the matter is well understood in perturbation 
theory (e.g., \citealt{Fry84,Goroff86,Bernardeau92,Jain94,Scoccimarro96,
Scoccimarro98,Scoccimarro99,Bernardeau02}). For Gaussian initial conditions, 
second-order perturbation theory predicts that the amplitude of 
the 3PCF $\xithree$ scales like the square of the amplitude of the two-point 
correlation function (2PCF) $\xi$ (\citealt{Peebles80}), and this scaling is
one of the fundamental tests of the Gaussian primordial fluctuations 
predicted by inflationary cosmology. Nonlinear gravitational 
evolution produces anisotropic, filamentary structures, so elongated
triangle configurations have higher amplitude in the three-point statistics.
Galaxy bias, a difference between the distributions of galaxies and 
matter, can alter three-point statistics, but it tends to boost
or suppress amplitudes for all triangle configurations equally, at least
on large scales. The triangle shape dependence of the 3PCF or bispectrum,
in combination with the 2PCF or power spectrum, therefore becomes a 
diagnostic for galaxy bias and the matter clustering amplitude 
(\citealt{Fry94}).

Early measurements of the 3PCF or bispectrum were based on angular
catalogs (e.g., \citealt{Peebles75,Fry80,Jing91,Fry94}). Galaxy redshift 
surveys like the Two-Degree Field Galaxy Redshift Survey (2dFGRS; 
\citealt{Colless01}) and the Sloan Digital Sky Survey 
(SDSS; \citealt{York00}), are now large enough to allow measurements 
of the redshift-space 3PCF or bispectrum with high
signal-to-noise ratio (e.g., \citealt{Verde02,Jing04,Kayo04}). 
However, this brings in the additional complication of distortion by the 
peculiar motions of galaxies. Even for the 2PCF or power spectrum, the effect
of nonlinear redshift-space distortions persist to remarkably large scales 
(\citealt{Cole94}). Distortions of the 3PCF or bispectrum are more
complex, and while some models of these distortions exist (e.g., 
\citealt{Scoccimarro01,Verde02}), it is not clear that they are accurate 
at the level of precision afforded by current data. An obvious way to 
circumvent redshift-space distortions is to project the 3PCF over the 
redshift direction, as is commonly done for the projected 2PCF $w_p(r_p)$ 
(\citealt{Davis83}). We note that projection of redshift surveys
is {\it not} the same as simply measuring from the parent angular catalog
--- the use of galaxy redshifts to obtain physical projected separations 
greatly reduces the noise and yields a quantity more closely related to the 
three-dimensional 3PCF. Projected 3PCFs have been measured for the Las 
Campanas Redshift Survey (LCRS) by \citet{Jing98} and for the 2dFGRS 
(\citealt{Jing04}) on relatively small scales. By comparing
the measurements with the predicted 3PCFs of matter in $N$-body
simulations, they find that the observed galaxy 3PCFs are lower than the 
predicted matter 3PCFs, indicating the need of galaxy bias to explain the 
data. 

Since the three-dimensional 3PCF is already a complicated object, projection 
may seem to forgo any chance of analytic treatment. However, I show here that 
the relation of the projected 3PCF to the (three-dimensional) bispectrum is 
relatively straightforward if the bispectrum is decomposed into Fourier 
moments (\S~2). My analysis is directly analogous to that of \citet{Szapudi04},
who introduces multipole expansion of the three-dimensional 3PCF and 
bispectrum and shows it to be useful in simplifying the relation between them, 
in characterizing the triangle configuration dependence, and in constraining 
galaxy bias. The spirit of this paper parallels that of \citet{Szapudi04}.
In \S~3 I show that bias effects remain simple for projected 2PCFs and 3PCFs 
on large scales, which makes it attractive to constrain galaxy bias using 
projected quantities. I briefly summarize my results in \S~4.  

\section{Fourier Expansion and Projected 3PCFs}

Before going into details, I emphasize that the meaning of projection in 
this paper has subtle differences from the usual angular correlation. 
Our goal is to describe the projection of the measured redshift-space 
correlation functions from a galaxy redshift survey and to use it to study 
problems like galaxy bias. Since galaxies with different properties cluster 
differently (e.g., red galaxies are in general more strongly clustered than 
blue galaxies), it is desirable to construct volume-limited samples that 
uniformly represent galaxies of a given type. By doing so, we can infer bias 
information as a function of galaxy type, rather than some average over all 
types of galaxies (weighted in a complicated way) like that from flux-limited 
galaxy samples. This approach helps the understanding of galaxy bias and 
aids comparisons with theoretical structure formation models. Redshift-space 
correlation functions can be measured by comparing
the galaxy distribution with a distribution of random points that has 
the same geometry and selection function as the galaxies (e.g.,  
\citealt{Landy93}). The projected correlation function can be formed
by integrating the redshift-space correlation function along the redshift
direction. The selection function does not enter the integration because 
its effect is included in the error budget of the measured redshift-space 
correlation function. In contrast, one needs to take into
account the selection function in modeling angular correlation measurements
of galaxy clustering. Nevertheless, our results in this paper can be 
easily generalized to include selection function to model angular 
clustering (see the discussion in \S~3 and \citealt{Fry99}). 

Since redshift-space distortions conserve numbers of pairs and triplets, 
2PCFs and 3PCFs projected from real-space three-dimensional correlation 
functions are identical to those from redshift-space three-dimensional 
correlation functions. This vastly simplifies theoretical predictions of 
projected 2PCFs and 3PCFs --- we only need to project the real-space 
three-dimensional correlation functions, without calculating redshift-space 
distortions. In practice, the radial extent of the projection may not be 
infinite, but it should be large enough to minimize any residual 
redshift-space distortions. We generally assume ideal infinite projections 
and plane-parallel geometry (the distant observer approximation) in our 
derivation. 

\subsection{Projected 2PCFs}

We start by reviewing the calculation of the projected 2PCF 
and its relation to the fluctuation power spectrum.

The (three-dimensional) 2PCF $\xi(\rr)$ is the Fourier transform 
of the fluctuation power spectrum $P(\kk)$,
\begin{equation}
\label{eqn:xi}
\xi(\rr)=\frac{1}{(2\pi)^3} \int d^3\kk\, P(\kk) e^{i\kk \bcdot \rr},
\end{equation}
where $\rr$ is the pair separation. Equation~(\ref{eqn:xi}) holds for
real-space quantities, which are isotropic, or for redshift-space quantities,
which are not. Both $\rr$ and $\kk$ can be decomposed into components 
perpendicular and parallel to the line of sight: $\rr=\rr_p+\rr_\parallel$ 
and $\kk=\kk_p+\kk_\parallel$. The projected 2PCF 
is obtained by integrating $\xi(\rr)$ along the line of sight,
\begin{equation}
\label{eqn:wp}
w_p(\rr_p)=\int_{-\infty}^{+\infty} dr_\parallel\, \xi(\rr_p+\rr_\parallel).
\end{equation}
After substituting equation~(\ref{eqn:xi}) into equation~(\ref{eqn:wp}) and 
putting variables into the form of perpendicular and parallel components, 
we have
\begin{eqnarray}
w_p(\rr_p) & = & \frac{1}{(2\pi)^3} \int d^2\kk_p\, e^{i\kk_p \bcdot \rr_p}
      \int_{-\infty}^{+\infty} dk_\parallel\, P(\kk_p+\kk_\parallel)
      \int_{-\infty}^{+\infty} dr_\parallel\, e^{ik_\parallel r_\parallel}
  \label{eqn:wp_a} \\
  & = & \frac{1}{(2\pi)^2} \int d^2\kk_p\, P(\kk_p) e^{i\kk_p \bcdot \rr_p}.
  \label{eqn:wp_b}
\end{eqnarray}
Equation~(\ref{eqn:wp_b}) follows from equation~(\ref{eqn:wp_a}) because
the rightmost integral in equation~(\ref{eqn:wp_a}) is just 
$2\pi$ times the Dirac $\delta$-function $\delta_D(k_\parallel)$. 
Equation~(\ref{eqn:wp_b}) states that the projected 2PCF is the 
two-dimensional Fourier transform of the power spectrum, and because
we have projected out redshift-space distortions, we can use the isotropic, 
real-space $P(k)$ in the integral. We can evaluate equation~(\ref{eqn:wp_b}) 
in polar coordinates. The angular part can be calculated by using the expansion 
$\exp(i\kk \bcdot \rr) = \sum_{n=-\infty}^{+\infty} J_n(kr)i^n
\exp[in(\phi-\Phi)]$ (plane waves in terms of cylindrical waves), 
where $\phi-\Phi$ is the angle between $\kk$ and $\rr$
and $J_n(x)$ is the Bessel function of integer order. Finally, 
equation~(\ref{eqn:wp_b}) reduces to a one-dimensional integral, 
\begin{equation}
\label{eqn:wpFT}
w_p(r_p)=\int_0^\infty \frac{k}{2\pi} dk\, P(k)J_0(kr_p).
\end{equation} 
This equation mimics the relation for the real-space, three-dimensional 2PCF 
which involves the spherical Bessel function $j_0(x)$. This kind of result 
can be found in papers that deal with projected observations, such as 
variants of Limber's equation (e.g., \citealt{Baugh93}). 

\subsection{Projected 3PCFs}

The projected 3PCFs can be derived in a similar way to the 2PCFs.

The three-dimensional 3PCF is the Fourier transform of the 
bispectrum $B(\kk_1,\kk_2,\kk_3)$,
\begin{equation}
\label{eqn:xithree_a}
\xithree(\rr_1,\rr_2,\rr_3)  =  \frac{1}{(2\pi)^6} 
  \int d^3\kk_1 d^3\kk_2 d^3\kk_3\, B(\kk_1,\kk_2,\kk_3)
  e^{i(\kk_1 \bcdot \rr_1 + \kk_2 \bcdot \rr_2 + \kk_3 \bcdot \rr_3)}
  \delta_D(\kk_1+\kk_2+\kk_3).
\end{equation}
The Dirac $\delta$-function selects wavevectors $(\kk_1,\kk_2,\kk_3)$ that 
form a triangle. From now on, we adopt the notation that 
a subscript of one number represents one side for a wavevector triangle, 
while it represents one vertex for a real-space triangle. Each side of 
a real-space triangle is denoted by subscript of two numbers, 
e.g., $\rr_{ij}=\rr_i-\rr_j$ ($i, j=1, 2, 3$).
Equation~(\ref{eqn:xithree_a}) reduces to
\begin{equation}
\label{eqn:xithree_b}
 \xithree(\rr_1,\rr_2,\rr_3) 
= \frac{1}{(2\pi)^6} \int d^3\kk_1 d^3\kk_2 B(\kk_1,\kk_2) 
  e^{i(\kk_1 \bcdot \rr_{13} + \kk_2 \bcdot \rr_{23})},
\end{equation}
where $\kk_3$ does not appear explicitly in the bispectrum because the
wavevector triangle is fully determined by $\kk_1$ and $\kk_2$ 
($\kk_1+\kk_2+\kk_3=0$).

Similar to the projected 2PCF, the projected 
3PCF for a projected triangle 
with $(\rr_{p1},\rr_{p2},\rr_{p3})$ as vertices can be obtained by
integrating $\xithree$ along the line of sight. One vertex 
(e.g., $\rr_{p3}$) can be fixed. We then have the following expression for 
the projected 3PCF,
\begin{equation}
\label{eqn:xithreeproj}
 \wpthree(\rr_{p1},\rr_{p2},\rr_{p3}) 
= \int_{-\infty}^{+\infty} dr_{\parallel 1}
  \int_{-\infty}^{+\infty} dr_{\parallel 2}\,
  \xithree(\rr_{p1}+\rr_{\parallel 1},
           \rr_{p2}+\rr_{\parallel 2},\rr_{p3}).
\end{equation}
Substituting equation~(\ref{eqn:xithree_b}) into the above, changing to 
cylindrical coordinates, and following the procedures used to go from 
equation~(\ref{eqn:wp_a}) to equation~(\ref{eqn:wp_b}), we can derive
the relation between the projected 3PCF
and the bispectrum 
\begin{equation}
\label{eqn:wpthreeFT}
 \wpthree(\rr_{p1},\rr_{p2},\rr_{p3}) 
= \frac{1}{(2\pi)^4} \int d^2\kk_{p1} d^2\kk_{p2}\, B(\kk_{p1},\kk_{p2}) 
 e^{i\kk_{p1} \bcdot \rr_{p13}} e^{i\kk_{p2} \bcdot \rr_{p23}}.
\end{equation} 
That is, the projected 3PCF is the two-dimensional Fourier transform of the 
bispectrum. Equation~(\ref{eqn:wpthreeFT}) is a four-dimensional integral, 
and it is clearly desirable to simplify the calculation, by reducing its 
dimensionality. Note that $B(\kk_{p1},\kk_{p2})$ is just the usual 
three-dimensional, real-space bispectrum evaluated at $\kk_1=\kk_{p1}$, 
$\kk_2=\kk_{p2}$, and $\kk_3=-(\kk_{p1}+\kk_{p2})$. 

\citet{Szapudi04} introduces multipole expansion of the three-dimensional 
three-point statistics and shows that the 3PCF can then be put in a simple 
form (see also \citealt{Verde00} for a similar expansion in the bispectrum). 
Consequently, the whole calculation becomes simple because only a few 
multipoles are needed for accurate convergence. In the same spirit as 
\citet{Szapudi04}, we introduce Fourier expansion of the projected three-point 
statistics. The bispectrum can be expanded as
\begin{equation}
B(k_1,k_2,\phi) = \sum_{n=-\infty}^{+\infty} B_n(k_1,k_2)e^{in\phi},
\end{equation}
where $\phi$ is the angle between $\kk_1$ and $\kk_2$,
and the coefficient $B_n(k_1,k_2)$ can be obtained through 
\begin{equation}
B_n(k_1,k_2)
=\frac{1}{2\pi} \int_0^{2\pi} d\phi\, B(k_1,k_2,\phi) e^{in\phi}
=\frac{1}{2\pi} \int_0^{2\pi} d\phi\, B(k_1,k_2,\phi) \cos n\phi.
\end{equation}
We have used the fact that the bispectrum has the symmetry 
$B(k_1,k_2,-\phi)=B(k_1,k_2,\phi)$, so $B_{-n}=B_n$
and the expansion is a cosine Fourier expansion.  

We rewrite equation~(\ref{eqn:wpthreeFT}) in polar variables, 
replace $B$ with its Fourier expansion, and expand the two exponentials
in the same way as we do in deriving equation~(\ref{eqn:wpFT}) (it is useful
to write the angle between two vectors as the difference of their polar 
angles). After integrating the angular part using the orthogonality of 
$\exp(in\phi)$, we find that
\begin{equation}
\wpthree(r_{p13},r_{p23},\Phi)
=\sum_{n=-\infty}^{+\infty} \wpthreen(r_{p13},r_{p23}) e^{in\Phi},
\end{equation}
where $\Phi$ is the angle between $\rr_{p13}$ and $\rr_{p23}$, and
the Fourier coefficient
\begin{equation}
\label{eqn:wpthreen}
\wpthreen(r_{p13},r_{p23})
=\int_{0}^{\infty} \frac{k_1}{2\pi} dk_1 
 \int_{0}^{\infty} \frac{k_2}{2\pi} dk_2\,
 (-1)^n B_n(k_1,k_2) J_n(k_1 r_{p13})J_n(k_2 r_{p23}).
\end{equation}
Equation~(\ref{eqn:wpthreen}) is in a form that resembles the
projected 2PCF in equation~(\ref{eqn:wpFT}). The essence here is
that, by Fourier expansion of the bispectrum $B$ and cylindrical wave 
expansion of the plane wave $\exp(i\kk_p\bcdot\rr_p)$, the angular dependence
is separated and integrated. There are also similarities between the 
above equation and equation (3) in \citet{Szapudi04} for the multipole 
coefficient of the three-dimensional 3PCF. We see that the calculation of the 
projected 3PCF from a given three-dimensional bispectrum requires essentially 
the same amount of work as the three-dimensional 3PCF (and a similar remark 
holds for 2PCF and the three-dimensional power spectrum). For 3PCFs, what we 
need to calculate are expansion coefficients, where only one-dimensional (for 
expansion coefficients of the bispectrum) and two-dimensional (for 
coefficients of the correlation function) integrals are involved.
I note that the same result (equation~[\ref{eqn:wpthreen}]) is also
obtained independently by I. Szapudi (2004, private communication).

If the effect of finite projection has to be taken into account, the Dirac 
$\delta$-function we use to derive equations~(\ref{eqn:wp_b}) and 
(\ref{eqn:wpthreeFT}) is replaced by 
$r_{\parallel,{\rm max}} j_0(k_\parallel r_{\parallel,{\rm max}}) /\pi$,
if the projection is performed for line-of-sight separations from
$-r_{\parallel,{\rm max}}$ to $r_{\parallel,{\rm max}}$.
Equations~(\ref{eqn:wp_b}) and (\ref{eqn:wpthreeFT}) become three- and 
six-dimensional integrals, respectively. The Fourier expansion can still
be adopted to reduce the calculation of the projected 3PCF to 
four-dimensional integrals.

As an example of the advantage of the Fourier expansion, Figure~1
shows the first few Fourier expansion coefficients of the matter bispectrum 
for $k_1=2k_2=0.05 h{\rm Mpc}^{-1}$ and the bispectrum recovered from using 
Fourier components up to $n=3,$ 5, 
and 10 in the weakly nonlinear regime. The matter bispectrum in this
regime can be written in the following form (e.g., \citealt{Fry84}),
\begin{equation}
\label{Bwnl}
B(k_1,k_2,\phi)=\left[\left(\frac{3}{2}-\frac{1}{2}\mu\right)
                     +\left(\frac{k_1}{k_2}+\frac{k_2}{k_1}\right)\cos\phi
                     +\frac{1}{2}(1-\mu)\cos 2\phi\right]P(k_1)P(k_2)
                     +{\rm perm.},
\end{equation}
where $P(k)$ is the linear matter power spectrum and $\mu$ reflects the weak 
dependence on cosmology ($\mu=3\Omega_m^{-1/140}/7\approx 3/7$ for
a spatially flat universe; \citealt{Kamionkowski99}, see also 
\citealt{Matsubara95}). The first term has Fourier components only up to 
$n=2$. The other permutations have higher frequency components because 
$k_3$, $\kk_1\bcdot\kk_3$, and $\kk_2 \bcdot \kk_3$ are all functions of 
$\cos\phi$. 

The results in Figure~1 are analogous to those of \citet{Szapudi04}, where 
the bispectrum is viewed as contributions by multipoles. In fact, the 
$n$th-order Legendre polynomial has Fourier components up to $n$. 
In Figure~1 we have adopted a linear power spectrum in the parameterization
of \citet{EBW92} with a primordial fluctuation power-law index $n_s=1$ and 
a shape parameter $\Gamma=0.21$. We see that only a few, low-frequency 
Fourier components are significant. For example, truncating the Fourier
series up to $n=5$ can recover the bispectrum with a fractional error 
of $\sim$1\% for $k_1=2k_2=0.05 h{\rm Mpc}^{-1}$. In the weakly nonlinear
regime, the galaxy bispectrum can be obtained by a linear combination of 
the matter bispectrum and the quantity 
$P(k_1)P(k_2)+P(k_2)P(k_3)+P(k_3)P(k_1)$ with combination coefficients
depending on galaxy bias (e.g., \citealt{Fry94}). In modeling galaxy 3PCFs, 
it is useful to expand this quantity in Fourier series, too.
Again, the first few Fourier coefficients are enough.  

\section{Reduced Projected 3PCFs and Galaxy Bias}

The reduced 3PCF in real space is defined as
\begin{equation}
\label{eqn:Qdef}
Q(\rr_1,\rr_2,\rr_3)=\frac{\xithree(\rr_1,\rr_2,\rr_3)}
{\xi(r_{12})\xi(r_{23})+\xi(r_{23})\xi(r_{31})+\xi(r_{31})\xi(r_{12})}.
\end{equation}
For galaxies, this quantity has contributions from both gravity and galaxy
bias (e.g., \citealt{Fry93,Fry94,Juszkiewicz95}). For Gaussian initial
conditions, in the weakly nonlinear 
regime, if the local galaxy overdensity 
$\delta_g$ is expanded in terms of the local matter overdensity $\delta_m$ 
as $\delta_g=\sum b_n \delta_m^n/n!$ (local bias model), the reduced 3PCFs 
of galaxies and matter have the relation
\begin{equation}
\label{eqn:Qbias}
Q_g(\rr_1,\rr_2,\rr_3) = \frac{1}{b} Q_m(\rr_1,\rr_2,\rr_3) + \frac{b_2}{b^2},
\end{equation}
where $b=b_1$ is the linear galaxy bias factor and $b_2$ is the lowest-order 
nonlinear bias factor. In redshift space, the 3PCF of galaxies, 
which is easy to measure, differs from the real-space one 
because of the additional contribution from peculiar motions of galaxies, 
and the above relation no longer holds. Constraints on galaxy bias from 
observations in redshift space therefore require accurate modeling of the 
peculiar motion of galaxies. Since projection along the line of sight 
essentially eliminates the effect of redshift-space distortions, it is 
interesting to see whether we can infer $b$ and $b_2$ using measurements of 
projected quantities. 

At first glance, it seems impossible to directly perform projections on 
both sides of equation~(\ref{eqn:Qbias}), since the $b_2/b^2$ term leads to
divergence. Furthermore, the projection of $Q$ from redshift space 
is not, in general, the same as from real space. We therefore project the
2PCFs and 3PCFs themselves, before taking ratios. To do so, we substitute 
the definition of $Q$ (equation~[\ref{eqn:Qdef}]) into 
equation~(\ref{eqn:Qbias}) and multiply both sides by 
[$\xi_g(r_{p12})\xi_g(r_{p23})+{\rm perm.}$]. On large scales, by using the 
relation $\xi_g=b^2\xi_m$, we find
\begin{equation}
\label{eqn:PCFgm}
\xithree_g(\rr_1,\rr_2,\rr_3)=b^3\xithree_m(\rr_1,\rr_2,\rr_3)
     +\frac{b_2}{b^2}\left[\xi_g(r_{12})\xi_g(r_{23})
     +\xi_g(r_{23})\xi_g(r_{31})+\xi_g(r_{31})\xi_g(r_{12})\right].
\end{equation}
Now we can perform projection on both sides by fixing $\rr_{p3}$, as we
do in equation~(\ref{eqn:xithreeproj}). After the projection, $\xithree$
becomes $\wpthree$ and it is easy to show that the product
$\xi_g(r_{ij})\xi_g(r_{jk})$ becomes $w_{p,g}(r_{pij}) w_{p,g}(r_{pjk})$.
We then divide both sides by
[$w_{p,g}(r_{p12}) w_{p,g}(r_{p23})+{\rm perm.}$] and use the linear bias 
relation $w_{p,g}=b^2w_{p,m}$ on large scales. The result is
just equation~(\ref{eqn:Qbias}) in terms of the reduced projected 3PCF 
$Q_p$, where $Q_p$ is formed as
\begin{equation}
\label{eqn:Qpdef}
Q_p(\rr_{p1},\rr_{p2},\rr_{p3}) = \frac{\wpthree(\rr_{p1},\rr_{p2},\rr_{p3})}
{w_p(r_{p12})w_p(r_{p23})+w_p(r_{p23})w_p(r_{p31})+w_p(r_{p31})w_p(r_{p12})}.
\end{equation}
So the relation between galaxy and matter reduced 3PCFs in 
equation~(\ref{eqn:Qbias}) still holds, 
\begin{equation}
\label{eqn:Qbiasp}
Q_{p,g}=\frac{1}{b}Q_{p,m}+\frac{b_2}{b^2}, 
\end{equation}
if the reduced projected 3PCF is properly defined. This kind of 
definition is also adopted by \citet{Jing98,Jing04}, who
point out that if $Q$ is a constant (which in general it is not) then $Q_p$ 
is also a constant and equal to $Q$. \citet{Jing98,Jing04} also 
measure the reduced projected 3PCF as a function of triangle shape for 
galaxies in the LCRS and the 2dFGRS.

Based on the above results, it is straightforward to infer bias from 
projected quantities. By projecting the measured redshift-space $\xi$ and 
$\xithree$, we form $Q_{p,g}$ for projected triangles that have 
two sides ($r_{p13}$ and $r_{p23}$) fixed but differ in the angle 
$\Phi$ between them. The dark matter $Q_{p,m}$ can be computed 
using the method introduced in \S~2 (for projecting $\xi$ and $\xithree$)
or can be measured from cosmological $N$-body simulations. The reduced 3PCFs
$Q_{p,g}(r_{p13},r_{p23},\Phi)$ and $Q_{p,m}(r_{p13},r_{p23},\Phi)$ 
can be expanded in either Fourier or multipole series, whichever one prefers. 
Their expansion coefficients satisfy $Q_{p,g,n}(r_{p13},r_{p23})
=Q_{p,m,n}(r_{p13},r_{p23})/b + \delta_{0n}b_2/b^2$, where $\delta_{0n}=1$
for $n=0$, and 0 otherwise. The nonlinear bias
factor only enters in the $n=0$ component (i.e., the DC component for
Fourier expansion or the monopole for multipole expansion). The relation
between coefficients of other components provide useful tests of bias
models and perturbation theory (see \S~4 of \citealt{Szapudi04} for a more 
detailed discussion, which also applies here). Alternatively, we can infer 
$b$ and $b_2$ from a two-parameter fit to $Q_{p,g}(r_{p13},r_{p23},\Phi)$, 
which is widely used (e.g., \citealt{Fry94,Feldman01,Scoccimarro01}). 

In order to make meaningful comparisons with predictions in the weakly 
nonlinear regime, all quantities should be measured on large scales. If the 
two sides ($r_{p13}$ and $r_{p23}$) of the projected triangle are large 
(e.g., greater than 10$h^{-1}{\rm Mpc}$) and differ substantially 
(e.g., $|r_{p13}-r_{p23}|>5h^{-1}{\rm Mpc}$), then the condition is guaranteed 
because any deprojected triangle lies in the weakly nonlinear regime
(e.g., $r>5h^{-1}{\rm Mpc}$). For configurations with $r_{p13}\approx r_{p23}$
(but still large), the strongly nonlinear regime affects the value of $Q_p$ at
$\Phi\sim 0$, where $r_{p12}$ approaches the size of the largest halos. Even 
for this extreme case, we can fit $Q_{p,g}$ by rejecting data points near 
$\Phi=0$. Since the strong nonlinear effect tends to add high-frequency 
components or higher order multipoles to $Q_p$, it is unlikely for it to 
have a large impact on bias parameters estimated from Fourier coefficients 
or multipole coefficients. The use of projected triangles with different 
values of $r_{p13}$ and $r_{p23}$ would lead to a consistency check of the 
inferred bias factors and increase the signal-to-noise ratio. 

As an illustration, we fit the projected 3PCFs measured by \citet{Jing04} for
galaxies in the 2dFGRS. \citet{Jing04} have measurements for two samples, 
one bright sample ($M_b\leq-19.66$) and one faint 
sample ($-19.66<M_b\leq 18.5$). They also have predictions of reduced 
projected 3PCFs of the matter for the concordance cosmological model, which
are measured in an $N$-body simulation. In their Figures~17 and 18, they 
compare the observed $Q_{p,g}$ and the predicted $Q_{p,m}$ for different 
triangle shapes, which provide the information we need to do the two-parameter
fit. A different parameterization is adopted by \citet{Jing04} to characterize 
a triangle, one parameter ($r_p$, the length of the shortest side) for the 
size and two parameters ($u$ and $v$)
for the shape, with the lengths of the three sides being $r_p$, $u r_p$, and 
$(u+v)r_p$. In their figures, the 3PCFs are plotted as a function of
$v$ (five equal linear bins in the range $0\leq v \leq 1$) for several 
combinations of $r_p$ and $u$. For the two-parameter fit, we choose the 
case with the largest value of $r_p$ they have, i.e., $3.25h^{-1}{\rm Mpc}$, 
and $u=2.09$. If the $v$ dependence is translated into an angular 
dependence, the five data points only cover $\Phi$ from $\sim 80^\circ$ to 
$\sim 180^\circ$, with the widths of the five angular bins being 
$12^\circ$, $13^\circ$, $15^\circ$, $19^\circ$, and $44^\circ$. We see that 
the last bin smears the angular dependence a lot. So when measuring the 3PCFs 
from galaxy clustering data, adopting the $(r_{p13},r_{p23},\Phi)$ 
parameterization and dividing $\Phi$ into narrow bins are probably more 
suitable to probe the angular dependence than the $(r_p,u,v)$ parameterization.
Since the errorbars on the measurements are large and the scales are not 
truly in the weakly nonlinear regime, we cannot obtain robust constraints 
on galaxy bias parameters. Nevertheless, application of the formalism gives 
bias factors for galaxies in the bright (faint) sample that are consistent 
with $b=1.8$ and $b_2=0$ ($b=1.1$ and $b_2=0$), somewhat higher $b$ than 
expected but not absurdly so.  

If the projection is not infinite, the kind of relation shown in 
equation~(\ref{eqn:Qbias}) also holds as long as we use the finite projected
correlation functions in the definition of $Q_p$ (equation~[\ref{eqn:Qpdef}]). 
Strictly speaking, in this case, the integration of the product of 2PCFs in 
the right side of equation~(\ref{eqn:PCFgm}) cannot be written as the product 
of $w_p$'s except for $\xi(r_{p13})\xi(r_{p23})$. However, the product of 
$w_p$'s should remain as a good approximation to the results. If one 
is not satisfied with the approximation, although it is good, one can always 
compute exact values of these integrations for $Q_{p,m}$ and form 
$Q_{p,g}$ from the observation in the {\it same} way. Based on 
equation~(\ref{eqn:PCFgm}), we still have the relation in 
equation~(\ref{eqn:Qbiasp}).

The result can also be generalized to a projected field, where 
projected correlations include the effect of the selection function 
(see \citealt{Fry99}).  Based on equation~(\ref{eqn:PCFgm}), it is easy 
to show that equation~(\ref{eqn:Qbiasp}) holds for $Q_p$ defined in terms of  
selection-function-weighted projected 2PCFs and 3PCFs. \citet{Fry99}
perform a systematic study of projection effects on the reduced 3PCFs.
They show that projections that are not deep enough would change the shape 
of the reduced 3PCF and thus bias the estimation of galaxy bias factors. 
However, this is based on the comparison with the {\it three-dimensional} 
reduced 3PCF of the matter. Our point here is that once the reduced 3PCF of 
the matter is calculated by taking account of the selection function, galaxy 
bias factors can be correctly inferred by comparing it with the observed 
reduced projected 3PCFs of galaxies. That is, we should always form 
$Q_{p,g}$ and $Q_{p,m}$ in the same way. The depth of the survey should 
be much larger than the extent of structures caused by galaxy peculiar 
velocities so that we can calculate 3PCFs of the matter in real space. 
\citet{Buchalter00} also investigate the reduced angular 3PCF. They 
concentrate on the explicit dependence of the galaxy 3PCF on cosmological 
parameters, the selection function, and bias, not on the relationship 
between the reduced 3PCF of galaxies and that of the matter as done here. 
Under certain assumptions, their results can be put into the form discussed
here. Focusing also on the explicit expression of the galaxy bispectrum,
\citet{Verde00} present a theoretical analysis of the projected bispectrum 
in spherical harmonics and discuss its application (note that they find that 
it is not encouraging to constrain bias parameters using projected galaxy 
catalogs).

\section{Summary and Discussion}

I investigate the projected 3PCF and find that it can be put in a
simple form if Fourier expansion is introduced. Each Fourier component
of the projected 3PCF is just a transform of the corresponding Fourier 
component of the bispectrum (equation~[\ref{eqn:wpthreen}]). In the weakly 
nonlinear regime, only the first few Fourier components are significant. The 
result is parallel to the multipole expansion proposed by \citet{Szapudi04} 
for the three-dimensional three-point statistics. Fourier expansion reduces 
the amount of work needed to compute the projected 3PCF and provides a 
convenient way to characterize its dependence on the triangle configuration. 
Projected 2PCFs and 3PCFs formed from redshift-space correlation functions 
are little affected by redshift-space distortions, which makes it promising 
to use them to measure galaxy bias factors. In the weakly nonlinear regime, 
I find that the relation between the reduced three-dimensional 3PCFs of 
galaxies and matter also holds for the reduced projected 3PCFs, if these are 
defined from the projected 3PCF and 2PCF in the same way. The linear bias 
factor $b$ and the first-order nonlinear bias factor $b_2$ thus can be 
inferred from the reduced projected 3PCFs of galaxies ($Q_{p,g}$) formed from 
redshift-space measurements, through the dependence of $Q_{p,g}$ on  
the angle between two sides of projected triangles. From the point of view 
of Fourier expansion, the nonlinear bias factor only affects the DC component 
of $Q_p$, and the linear bias factor is given by the ratio of coefficients of 
other Fourier components of $Q_{p,g}$ and $Q_{p,m}$.  

The method of measuring galaxy bias factors from projected 3PCFs can be 
directly applied to the data from contemporary galaxy redshift surveys 
(e.g., 2dFGRS and SDSS). The projected 3PCFs of galaxies should be measured 
on large scales, i.e., in the weakly nonlinear regime, where the bias relation 
(equation~[\ref{eqn:Qbiasp}]) predicted by the local bias model applies. 
The projected 3PCFs of matter can be calculated using the technique 
introduced in \S~2, or they can be measured from cosmological $N$-body 
simulations. The inferred bias factors of galaxies would help to 
constrain cosmological parameters, such as the amplitude of the matter
fluctuation power spectrum.

Theoretically, the framework of the halo occupation distribution is often 
adopted to model galaxy clustering by linking galaxies to dark matter halos
(for the three-dimensional 3PCF modeling within this framework, see, e.g., 
\citealt{Takada03,Wang04}). With the help of this framework, even in the 
strongly non-linear regime, galaxy clustering would bring in additional 
constraining powers on cosmological parameters as well as on the relation 
between galaxies and matter. In any regime, projected 3PCFs, little affected 
by redshift-space distortions, are useful statistics. On very small scales 
where the three galaxies of each triplet come from the same dark matter 
halo, 3PCFs probe the shape of halos and the relative distribution of 
galaxies and matter inside halos. The projected 3PCF in this regime can be 
obtained by projecting the real-space 3PCF, which is easy to calculate 
(\citealt{Takada03}). On intermediate or large scales where the galaxies of 
each triplet are from two or three different halos, the calculation of the 
3PCF is simpler in Fourier space than in real space. Fourier expansion 
would be a useful technique to calculate the projected correlation 
functions on these scales. 

\acknowledgments
I am grateful to David Weinberg for his encouragement and advice on this 
work and for his detailed and valuable comments that greatly improved the 
paper. I thank Istv\'an Szapudi and Jun Pan for useful suggestions and 
comments. This work was supported by a Presidential Fellowship from the 
Graduate School of the Ohio State University and by NSF grant AST 04-07125.

\begin{figure}[h]
\plotone{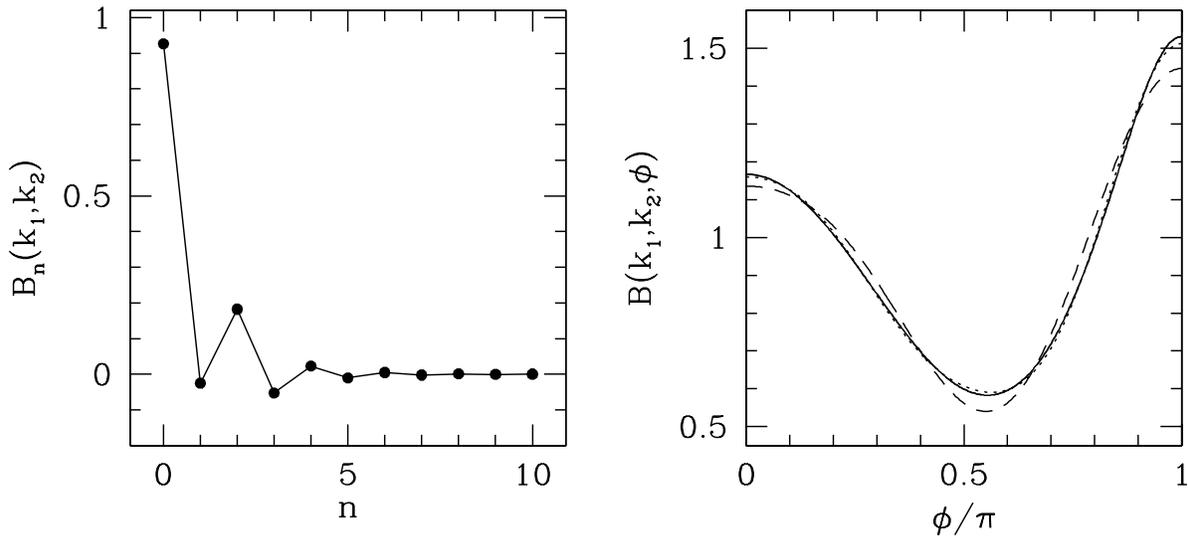}
\caption[]{Weakly nonlinear matter bispectrum and Fourier expansion 
coefficients for $k_1=2k_2=0.05 h^{-1}{\rm Mpc}$. The left panel shows 
the Fourier coefficients $B_n$, which become virtually zero for $n>5$. 
The right panel shows the bispectrum $B$ ({\it solid curve}) as a function of 
the angle $\phi$ between $\kk_1$ and $\kk_2$ and bispectra reconstructed 
from the first few Fourier components. The bispectra from the Fourier series 
up to $n=3$ ({\it dashed curve}), 5 ({\it dotted curve}), and 10 ({\it 
dot-dashed curve}, indistinguishable from the solid curve) have maximum 
fractional errors of $\sim$7\%, $\sim$1.5\%, and $\sim$0.04\%, respectively. 
A cold dark matter linear power spectrum is assumed (see the text). Both $B$ 
and $B_n$ are normalized by dividing $3P(k_1)P(k_2)$ for display.
}
\end{figure}

\end{document}